# *In Vivo* Renal Clearance, Biodistribution, Toxicity of Gold nanoclusters

*Xiao-Dong Zhang,\* Di Wu, Xiu Shen, Pei-Xun Liu, Fei-Yue Fan, and Sai-Jun Fan*

Institute of Radiation Medicine and Tianjin Key Laboratory of Molecular Nuclear Medicine, Chinese Academy of Medical Sciences and Peking Union Medical College
Xiaodongzhang@tju.edu.cn.

ABSTRACT: Gold nanoparticles have shown great prospective in cancer diagnosis and therapy, but they can not be metabolized and prefer to accumulate in liver and spleen due to their large size. The gold nanoclusters with small size can penetrate kidney tissue and have promise to decrease *in vivo* toxicity by renal clearance. In this work, we explore the *in vivo* renal clearance, biodistribution, and toxicity responses of the BSA- and GSH-protected gold nanoclusters for 24 hours and 28 days. The BSA-protected gold nanoclusters have low-efficient renal clearance and only 1% of gold can be cleared, but the GSH-protected gold nanoclusters have high-efficient renal clearance and 36 % of gold can be cleared after 24 hours. The biodistribution further reveals that 94% of gold can be metabolized for the GSH-protected nanoclusters, but only less than 5% of gold can be metabolized for the BSA-protected nanoclusters after 28 days. Both of the GSH- and BSA-protected gold nanoclusters cause acute infection, inflammation, and kidney function damage after 24 hours, but these toxicity responses for the GSH-protected gold nanoclusters can be eliminated after 28 days. Immune system can also be affected by the two kinds of gold nanoclusters, but the immune response for the GSH-protected gold nanoclusters can also be recovered after 28 days. These findings show that the GSH-protected gold nanoclusters have small size and can be metabolized by renal clearance and thus the toxicity can be significantly decreased. The BSA- protected gold nanoclusters, however, can form large compounds and further accumulate in liver and spleen which can cause irreparable toxicity response. Therefore, the GSH-protected gold nanoclusters have great potential for *in vivo* imaging and therapy, and the BSA-protected gold nanoclusters can be used as the agent of liver cancer therapy.

## 1. Introduction

Gold-based nanomaterials have been focused owing to their distinct optical properties and potential applications in photothermal therapy, radiation therapy, and contrast agent [1, 2, 3, 4, 5]. Therefore, *in vivo* toxicity of gold nanoparticles (NPs) has received wide interest [6, 7]. It has been found that the *in vivo* toxicity of gold NPs is directly related to the size, shape and surface coating, exposure dose, and administration routes [7, 8, 9, 10, 11, 12, 13, 14]. However, it is widely conceived that gold NPs have high-level accumulation in the liver and spleen, and these accumulations can induce the gene changes and liver necrosis [15, 16]. Therefore, it is desirable to find the metabolizable gold NPs to decrease the *in vivo* toxicity.

It has been demonstrated that the size plays a dominant role in metabolism of NPs. Small semiconductor NPs, namely quantum dots (QDs) can be cleared by renal filtration and urinary excretion [14]. It is suggested that QDs smaller than 5.5 nm can be rapidly and efficiently metabolized by renal clearance, and QDs larger than 15 nm can prevent the renal excretion and can be accumulated in the liver and spleen [16, 17, 18]. Indeed, in another independent work, it has also been found that 4-6 nm ZnS and CdSe QDs can be cleared by kidney, and only small amounts QDs can be found in the spleen, kidney, and bone marrow [14]. However, in the practical *in vivo* experiment, the metabolism of gold NPs is also closely related to surface chemistry. Naked gold NPs without coatings can cause serious aggregation and NPs can accumulate in liver. For example, all sizes of naked gold particles (5-100 nm) can cause high organ accumulation, and the liver and spleen are dominant targeting organs [19, 20, 21, 22]. Even using the coating, physiological environment may also cause the degeneration of coating and induce the failure of designing. A case in point is that the polyethylene glycol (PEG) coating can decrease the zeta potential of gold NPs and further prevent the aggregation partially in the blood, but these coating still can not decrease the liver and spleen accumulation and improve the metabolism of gold NPs. It has been found 5 nm PEG-coated gold NPs have long time circulation in blood [15, 23], and high accumulation in the liver [16, 21, 24]. Thus, it is necessary to explore the renal clearance of smaller and more stable gold nanomaterials and decrease the side effects.

Compared with traditional NPs, the size of gold nanoclusters (NCs) is smaller. Typical thiol-protected $Au_{25}$ NCs with the size of 1-2 nm will have stronger permeability and petention effect (EPR) than larger size particles. Besides, gold NCs have high luminescence efficiency and *in vivo* imaging can be achieved by fluorescence [25, 26, 27, 28, 29, 30, 31, 32] instead of scattering of NPs [1] or two-photon luminescence of nanorod [33, 34, 35]. On the one hand, it is necessary to explore the renal clearance and metabolism of these clusters in order to decrease the accumulation of liver and spleen. On the other hand, it is desirable to illustrate the *in vivo* toxicity of $Au_{25}$ luminescence NCs, which have important indications to both of the *in vivo* imaging and therapy. Herein, we perform the *in vivo* renal clearance and toxicity of the GSH- and BSA-protected $Au_{25}$ NCs by evaluating the biodistribution, immune response, hematology, and biochemistry.

**2. Materials and methods**
*2.1. Fabrication*

The BSA- and GSH-protected $Au_{25}$ NCs were synthesized following the recent reported procedures [26, 36]. Briefly, for the BSA-protected $Au_{25}$ NCs, $HAuCl_4 \cdot 3H_2O$ solution (17 mL, 5 mM, 37 °C) was added to the BSA solution (5 mL, 50 mg/mL, 37 °C), then, NaOH solution (0.5 mL, 1 M) was introduced, and the mixture was incubated at 37 °C for 12 h. The red photoluminescence of the BSA-protected $Au_{25}$ NCs was located at 650 nm. For the GSH-protected $Au_{25}$ NCs, reduced GSH (20 mM) was added into the $HAuCl_4 \cdot 3H_2O$ solution (8 mL, 5 mM) and an aqueous solution of $NaBH_4$, cooled at 0 °C, was injected rapidly into this mixture under vigorous stirring and ligand etching by adding GSH again. The photoluminescence was measured by using a Hitachi F4500 fluorescence spectrophotometer with a 100 W xenon lamp as the excitation source, and the excitation and emission wavelength are 480 nm and 650 nm, respectively. The size and morphology of the $Au_{25}$ NCs were analyzed by using a Hitachi HF-2000 field emission high-resolution TEM operating at 200 kV.

*2.2. Animal injections and sample collection*

Animals were purchased, maintained, and handled with protocols approved by the Institute of Radiation Medicine, Chinese Academy of Medical Sciences (IRM, CAMS). Female mice were obtained from IRM laboratories at 11 weeks of age and were housed 2 per cage in a 12 h/12 h light/dark cycle, and were given food and water ad libitum. Mice were randomly divided into three groups (six in each group): control group, the BSA- and GSH-protected $Au_{25}$ NCs treated groups, respectively. 151 µg/ml BSA- and GSH-protected $Au_{25}$ NCs were used for the animal experiment using intraperitoneal injection, and concentration was up to 7550 µg/kg in the mice. Animals were injected with 100 µL of control, the BSA-, and GSH-protected $Au_{25}$ NCs solution. At every day time points after injection, animals were weighed and assessed for behavioral changes. Using a standard saphenous vein blood collection technique, blood was drawn for hematology analysis (potassium EDTA collection tube). The analysis of standard hematological and biochemical examination was performed. For blood analysis, 200 µL of blood was collected from mice and separated by centrifugation into cellular and plasma fractions. Upon the completion of the last time point, mice were sacrificed by isoflurane anesthetic and angio catheter exsanguination. Major organs from those mice were harvested, fixed in 4% neutral buffered formalin, processed routinely into paraffin, stained with hematoxylin and eosin (H&E) and pathology are examined by a digital microscope.

*2.3. Biodistribution and TEM observation*

The organs and original solutions of the BSA- and GSH-protected $Au_{25}$ NCs treated mice were digested by using a microwave system CEM Mars 5 (CEM, Kamp Lintfort, Germany). The Au content was measured with an inductively coupled plasma mass spectrometer (ICP-MS, type Agilent 7500 CE, Agilent Technologies, Waldbronn, Germany). Blood cells were obtained after tail vein injection of 24 hours. For the TEM analysis, the solution was centrifuged and fixed with 2.5% glutaraldehyde in 0.03 M potassium phosphate buffer, pH 7.4. The cells were postfixed with 1% osmium tetroxide in 0.1 M sodium cacodylate buffer and 0.5% uranyl acetate in 0.05 M maleate buffer, which were then dehydrated in a graded series of ethanol and embedded in Epon. Ultrathin sections were cut and transferred on 200-mesh uncoated copper grids, stained with uranyl acetate, counter-stained with lead citrate.

## 3. Results and discussions

*3.1. Synthesis and characterization of $Au_{25}$ NCs*

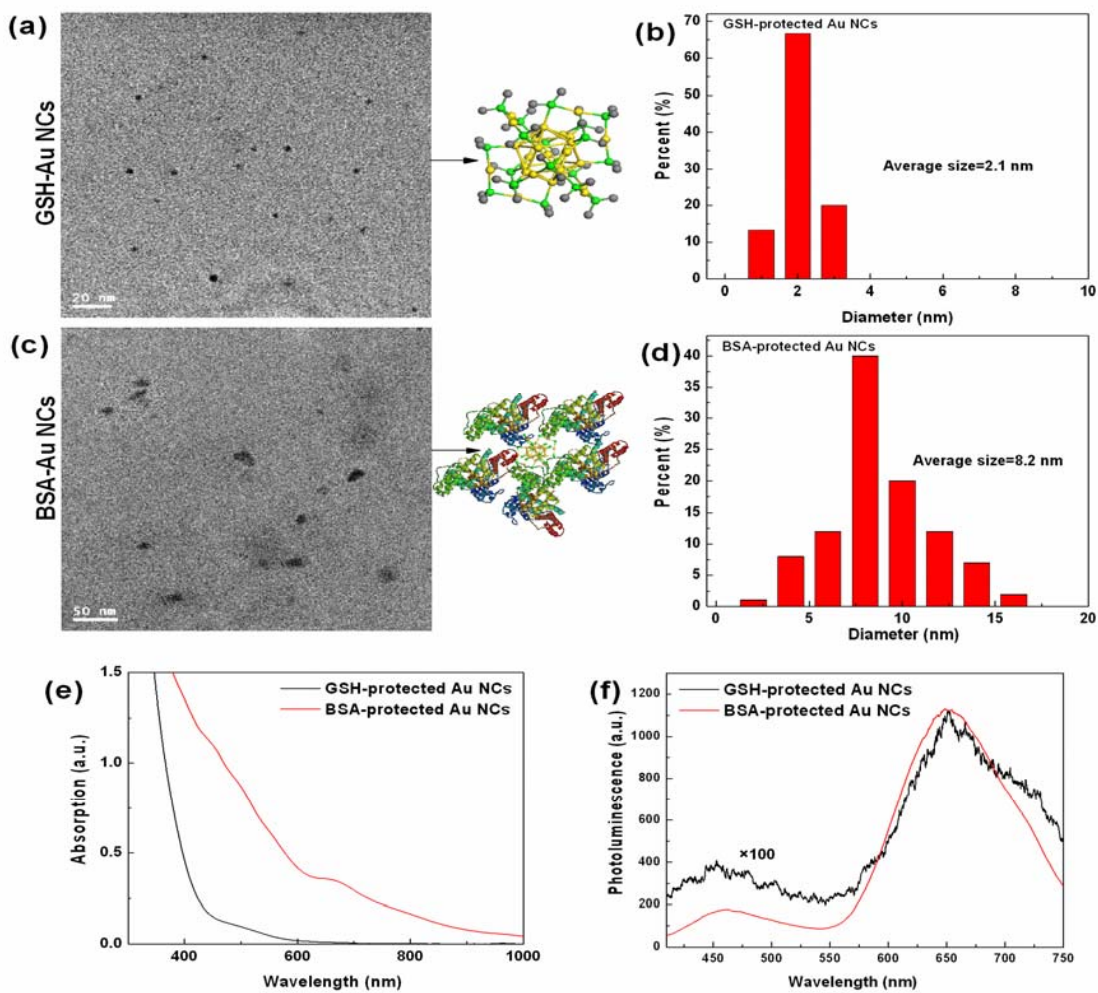

**Figure 1.** (a) The TEM images and (b) size distribution of the GSH-protected $Au_{25}$ NCs. (c) The TEM images and (d) size distribution of the BSA-protected $Au_{25}$ NCs. (e) Optical absorption and (f) photoluminescence of the GSH- and BSA-protected $Au_{25}$ NCs. It can be seen that the emission centre is located at 650 nm.

The size distribution and photoluminescence of the GSH- and BSA-protected $Au_{25}$ NCs are characterized by TEM and spectrophotometer in Fig.1. The average size of the monodisperse

GSH-protected Au$_{25}$ NCs is 2.1 nm. The surface of the BSA-protected Au$_{25}$ NCs has lots of BSA, which induces the increase of size up to 8.2 nm, similar to the recent reported results of 8.5 nm [31]. The large size of the BSA-protected Au$_{25}$ NCs is not difficult to understand, and the BSA contains lots of SH molecules which are easy to form large coating by Au-S interaction. The optical absorption of protected Au$_{25}$ NCs has been performed in Fig.1 (e). For the BSA-protected Au$_{25}$ NCs, the 420 and 650 nm absorption bands have been observed, which are due to the optical absorptions of the BSA-protected Au$_{25}$ NCs. The optical absorption of the GSH-protected Au$_{25}$ NCs presents 450 nm absorption band and the 650 nm band is too weak to observe. The optical absorption induces the luminescence of Au$_{25}$ NCs, and the photoluminescences of the two NCs are presented in Fig.1 (f). The red emission at 650 nm represents the direct optical transition and the emergence of thiol protected Au$_{25}$ NCs. Quantum efficiency of the BSA-protected Au$_{25}$ NCs is 0.06 while it is 0.001 for the GSH-protected Au$_{25}$ NCs, which are related to the ligand molecule [37]. The metal doping or binding with hole-rich molecule can enhance the emission [38, 39, 40].

*3.2. Blood plasma stability*

When the Au$_{25}$ NCs are injected, the NCs will firstly interact with blood plasma. To investigate human plasma stability and interaction with Au$_{25}$ NCs, Fig.2 (a) and (b) present the time-dependent optical absorption of the GSH- and BSA-protected Au$_{25}$ NCs in human blood plasma. It is necessary to point out that Au$_{25}$ NCs have not presented any surface plasmon resonance (SPR) absorption in Fig.1 (e), which is the feature of large size gold NPs. However, after dilution of human plasma, the GSH- and BSA-protected Au$_{25}$ NCs show the SPR absorption. The slight redshift of SPR has been observed in the two protected Au$_{25}$ NCs. which indicate that the Au$_{25}$ NCs have slightly aggregated in solution, and NCs may transform into NPs gradually. The intensity of SPR is gradually decreased for the two kinds of protected NCs. The decreased SPR is related to the structure and optical properties of gold, which indicate the NCs have combined with the proteins in blood plasma and form the larger compounds. The BSA is macromolecule and it can cover the optical absorption of NCs. It is well known that the naked and citrate coated gold NPs have high zeta potential, and it prefers to aggregation in physiological environment. The BSA- and GSH-protected Au$_{25}$ NCs have decreased the surface charge and activity, and it is helpful to improve the stability of NCs. To directly observe the NCs distribution in blood plasma, we performed the TEM observations.

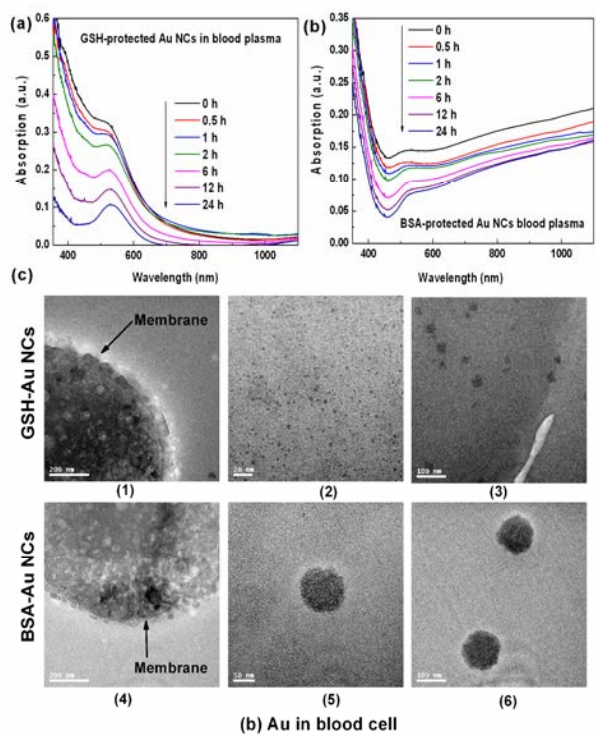

**Figure 2.** Stability of (a) GSH- and (b) BSA-protected Au$_{25}$ NCs in the human blood plasma has been tested using time-dependent optical absorption. (c) The TEM observations of the GSH- and BSA-protected Au$_{25}$ NCs in blood cell of mice.

As presented in Fig. 2 (c), the TEM images of the GSH- and BSA-protected Au$_{25}$ NCs in blood give microscopic insight for the distribution of NCs. Fig.2 (c1-c3) and (c4-c6) present the image of the GSH-protected Au$_{25}$ NCs and BSA-protected Au$_{25}$ NCs at different scales, respectively. For the GSH-protected Au$_{25}$ NCs after 24 hours interaction, it has been found the partial NCs have aggregated slightly and the average size is in the range of 5-30 nm in Fig.2 (c3). The NPs can be formed by several clusters, which can be widely observed. However, we also find lots of monodisperse NCs in Fig. 2(c2). We can conclude that some NCs have transformed to NPs, and the complex compounds including the biomolecule and gold can be found in the blood. However, the BSA-protected Au$_{25}$ NCs can form the large size NPs, and the average size is in the range of 40-80 nm. These results are in good agreement with the results of blood plasma stability in Fig.2 (a) and (b). These results are not surprising, and the PEG- and BSA-coated $Au_{25}$ NCs are easy to suffer aggregation in the solution physiological environment after long time exposure. These coating can reduce the zeta potential and surface activity, but blood protein includes lots of SH- and NH- group and can degenerate these coating. Therefore, gold can hybrid with some macromolecule proteins by Au-S and $Au-NH_3$ binding interaction which seem similar to surface coating [41, 42]. It can be found that the GSH-protected $Au_{25}$ NCs have better stability than the BSA-protected $Au_{25}$ NCs. Next, we will move to the renal clearance and metabolism of these NCs in in vivo experiment.

3.3. In vivo biodistribution and renal clearance

Fig.3 (a) and (b) presents in vivo biodistribution of the GSH-and BSA-protected $Au_{25}$ NCs after 24 hours and 28 days, because it is widely reported that gold NPs have long blood circulation time and bioaccumulation in the liver and spleen [15, 20, 43, 44, 45, 46, 47, 48]. For 24 hours distribution, liver and spleen have high content. The GSH-protected $Au_{25}$ NCs prefer to stay in spleen and the BSA-protected $Au_{25}$ NCs have high concentration in liver. After 28 days, biodistribution of the two clusters are quite different. The BSA-protected $Au_{25}$ NCs have very high concentrations in the liver (7295 ng/g) and spleen (6982 ng/g), but the GSH-protected $Au_{25}$ NCs have low concentrations. Recalling biodistribution of gold particles of 10-100 nm [20, 43], it can be found that Au concentration from the BSA-protected $Au_{25}$ NCs in the liver and spleen is 2 times higher than the previous results, but the Au concentration from the GSH-protected $Au_{25}$ NCs is 10 times lower than widely available results in liver [20, 43]. In addition, the BSA-protected $Au_{25}$ NCs can be found in the lung and kidney with Au concentration of 979 and 1206 ng/g, respectively, but the GSH-protected $Au_{25}$ NCs have only 114 and 224 ng/g in lung and kidney, respectively. Only in heart, the BSA-protected $Au_{25}$ NCs have lower concentration than the GSH-protected $Au_{25}$ NCs. The tiny Au content can be found in brain, which is not surprising because even 15 nm gold particles can also pass blood–brain barrier [43]. Both of the BSA and GSH- protected $Au_{25}$ NCs have high concentration in reproductive system. The BSA-protected $Au_{25}$ NCs have extremely high bioaccumulation in liver and spleen, but the Au concentration for the GSH-protected $Au_{25}$ NCs is sharply decreased. It is widely believed that the BSA-coated gold NPs are biocompatible materials in in vitro because BSA coating can modify surface chemistry and reduce toxicity [26, 49], but the present experiment results clearly show that the in vivo toxicity is quite different from the in vitro toxicity, and the metabolism of the BSA-protected $Au_{25}$ NCs is not good. However, the GSH- protected $Au_{25}$ NCs are very interesting, which show self-clearance function and can be metabolized partially.

To confirm these results and investigate the renal clearance, we also measure the Au concentration and content in urine. Fig.3 (c) shows the time-dependent Au concentration of the GSH and BSA-protected $Au_{25}$ NCs in the urine. For the GSH-protected $Au_{25}$ NCs, Au concentration is as high as 226 ng/g after 2 hours injection. With increasing time, the Au concentration gradually decreases to 38 ng/g. As a contrast, the Au concentration is only 1-4 ng/g for the BSA-protected $Au_{25}$ NCs at these time range. Moreover, Fig.3 (d) shows the time-dependent Au content of the GSH- and BSA-protected $Au_{25}$ NCs in the urine, which represents the metabolized Au amount by renal clearance. It can be seen that the total Au is 5568 ng for the GSH-protected $Au_{25}$ NCs and 227 ng for the BSA-protected $Au_{25}$ NCs after 24 hours metabolism, which indicate that 36 % of the GSH-protected $Au_{25}$ NCs and only 1% of the BSA-protected $Au_{25}$ NCs can be excreted by urine. After 28 days treatment, only 6% of Au can be found in the GSH-protected $Au_{25}$ NCs treated mice, but more than 95% of Au can be found in the BSA-protected $Au_{25}$ NCs treated mice. These results clearly show the GSH-protected $Au_{25}$ NCs can be metabolized by renal clearance. On the one hand, the average size of the GSH-protected $Au_{25}$ NCs is as small as 2.1 nm, and it can penetrate the kidney tissue to metabolize. On the other hand, the GSH-protected $Au_{25}$ NCs can slightly aggregate up to 5-30 nm gradually in the blood plasma with increasing time, but the metabolism rate of NCs can reach the maximum after 2 hours injection. However, the average size of the BSA-protected $Au_{25}$ NCs is about 8.2 nm. Furthermore, it can form NPs larger than 40 nm in blood, which prevent metabolism. It has been proposed that the NPs larger than 15 nm can not be metabolized, and the 5 nm NPs can be cleared by renal excretion [14]. Similar findings have been reported in the small ZnS QDs, and it was found that the QDs coated with the organic molecule mercaptoundecanoic acid and BSA were cleared from plasma with a clearance of 0.59-1.23 mL min$^{-1}$kg$^{-1}$ [17]. Small clusters may be cleared and metabolized by kidney.

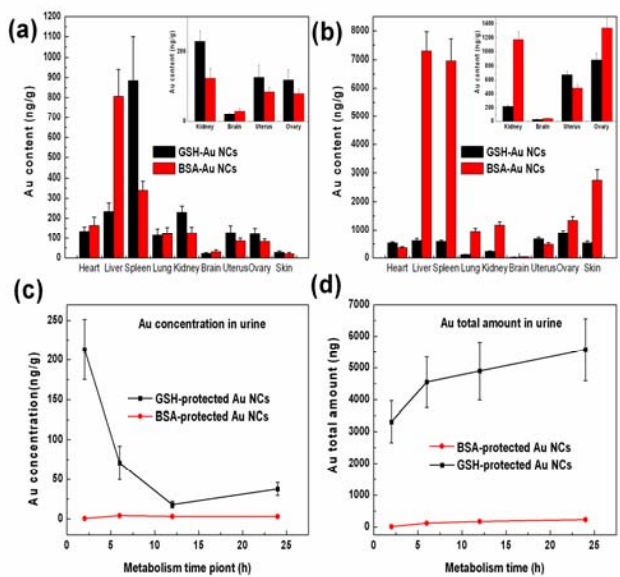

**Figure 3.** Biodistribution of the GSH- and BSA-protected Au$_{25}$ NCs treated mice at (a) 24 hours and (b) 48 hours. (c) Time dependent Au concentration (clearance rate) and (d) Au content (total metabolism) in 24 hours. The BSA-protected Au$_{25}$ NCs have 10 times higher distribution in the liver and spleen than the GSH-protected Au$_{25}$ NCs. Each point represents the mean ± standard deviation. Bars represent mean ± standard deviation.

*3.4. Body weight, immune response, and pathology*

Fig.4 (a) gives the body weight of mice treated by the two NCs. It can be seen that both of the BSA- and GSH- protected Au$_{25}$ NCs at the dose of 7550 μg/kg don't cause mortality compared to the control group within a 1-day observation period. During the study period, the treatment with Au$_{25}$ NCs for 28 days doesn't cause obvious adverse effects on growth, and no significant statistical differences are observed in body weight between the Au$_{25}$ NCs-treated mice and control mice. Further, no abnormal clinical signs and behaviors are detected in both control and treated groups. Necropsy of the mice at the end of the experiment doesn't show any macroscopic changes of the organs in these groups. To further investigate the immune reaction of organ, we give the organ indexes of thymus and spleen for 24 hours treated mice in Fig.4 (b) and for 28 days treated mice in Fig. 4 (c). To explicitly examine the grade of changes caused by malities, spleen and thymus indexes ($S_x$) can be defined as:

$$S_x = \frac{\text{Weight of experimental organ }(mg)}{\text{Weight of experimental animal }(g)}$$

For 24 hours treated mice, spleen index has no significant change. The thymus index, however, has significantly increased to 4.2 for the two kinds of protected Au$_{25}$ NCs. For 28 days treated mice, the average values of spleen and thymus index in control group are 3.3 and 2.3, respectively. The spleen indexes of the mice treated by the GSH- and BSA-protected Au$_{25}$ NCs have no significant changes at the concentration of 7550 μg/kg. However, the thymus index of the BSA-protected Au$_{25}$ NCs treated mice increases to 3.7, and there are significantly statistical differences between the treated groups and control group. The high dose of Au induces the acute immune responses after 24 hours treatment, and these responses of the GSH-protected Au$_{25}$ NCs are recovered after 28 days, but that of the BSA-protected have not recovered. The BSA-protected Au$_{25}$ NCs have induced more obvious immune response than that of the GSH-protected Au$_{25}$ NCs. We next check the pathological results of organ pathology by using immunohistochemistry in Fig.5. It can be seen that the two NCs treated heart, spleen, and kidney have not caused appreciable pathological lesion. However, the local liver necrosis has been observed in the BSA-protected Au$_{25}$ NCs, which indicate that liver suffers the unrecoverable damage. To quantify the toxicity of the two NCs, the next important step is assessment of standard hematology and biochemistry [50].

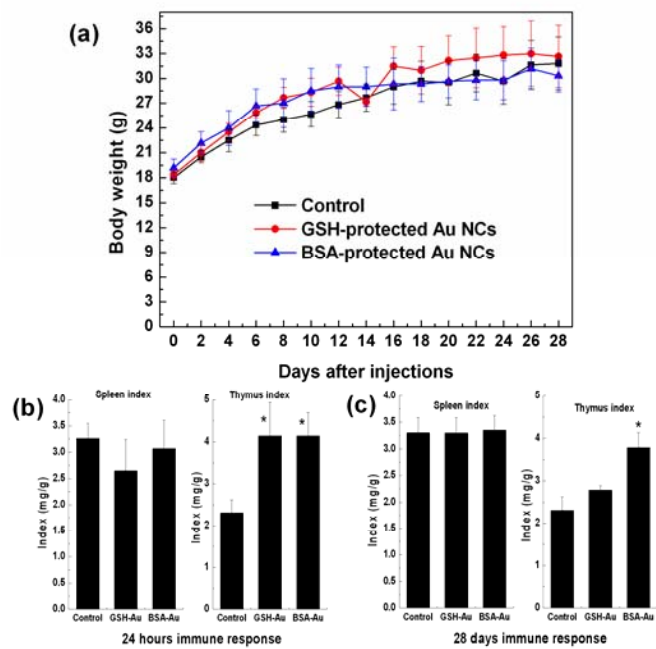

**Figure 4.** (a) Body weights of the GSH- and BSA-protected $Au_{25}$ NCs treated mice. Spleen index and thymus index of mice after (b) 24 hours and (c) 28 days treatments. Bars represent mean ± standard deviation. Data is analyzed by T student test, and star represents significant difference from the control group ($p < 0.05$).

*3.5. Hematology and biochemistry*

We select standard hematology markers for analysis, such as white blood cell (WBC), red blood cell (RBC), hematocrit (HCT), mean corpuscular volume (MCV), hemoglobin (HGB), platelet (PLT), mean corpuscular hemoglobin (MCH), and mean corpuscular hemoglobin concentration (MCHC). Hematology results for the GSH- and BSA-protected $Au_{25}$ NCs are presented in Figure 6. For 24 hours treated mice, the WBC and RBC in mice treated with the BSA-protected $Au_{25}$ NCs increase significantly. Similarly, the MCH and MCHC from mice treated with the two NCs also increase significantly. After 28 days, the WBC and HGB still increase for the BSA-protected $Au_{25}$ NCs treated mice. Other parameters, such as HCT, MCV, and PLT have no significant difference, which indicates tiny damage to the mice. However, all parameters have recovered for the GSH-protected $Au_{25}$ NCs treated mice after 28 days. The WBC is sensitive to the physiological response in mice. The rise of WBC in mice treated with the BSA- protected $Au_{25}$ NCs indicates an inflammatory response. After 28 days, the BSA-protected $Au_{25}$ NCs still show toxicity.

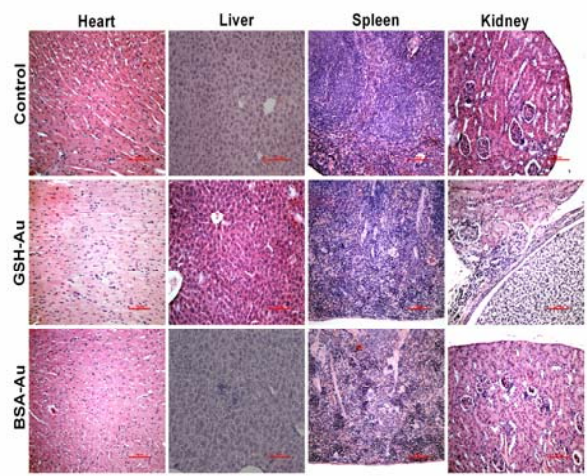

**Figure 5.** Pathological results from the heart, liver, spleen, and kidney of the GSH- and BSA-protected $Au_{25}$ NCs after 28 days. Appreciable pathological changes have not been found in all these organs expect for liver in the BSA-protected $Au_{25}$ NCs.

Furthermore, we present the biochemistry results of the GSH- and BSA-protected $Au_{25}$ NCs in Fig.7 including (a) Alanine transaminase (ALT), (b) Aspartate transaminase (AST), (c) total protein (TP), (d) albumin (ALB), (e) blood urea nitrogen (BUN), (f) creatinine (CREA), (g) globulin (GLOB), and (h) total bilirubin (TBIL). We emphasize the ALT, AST, and CREA, because they are closely related to the liver and kidney function of mice. After 24 hours treatment, it can be found that the CREA increases for the two kinds of $Au_{25}$ NCs treated mice, and the AST and BUN increase for the GSH-protected $Au_{25}$ NCs treated mice. Besides, the GLOB also increases for the two kinds of NCs treated mice. Both of the GHS- and BSA-protected $Au_{25}$ NCs have caused the acute damage for liver and kidney after 24 hours. After 28 days, the CREA for the BSA-protected $Au_{25}$ treated mice decreases and indicates that kidney function can not be recovered. However, all parameters for the GSH-protected $Au_{25}$ NCs treated mice have recovered.

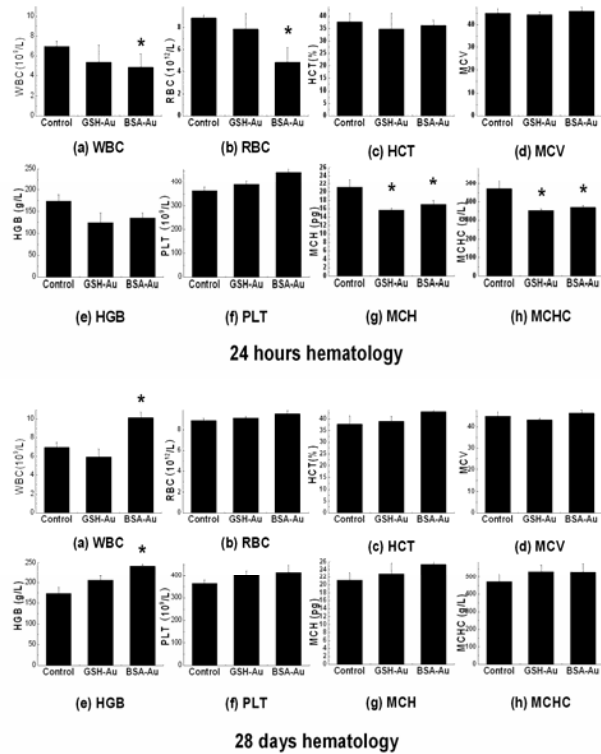

**Figure 6.** Hematology results of the GSH- and BSA-protected $Au_{25}$ NCs treated mice after 24 hours and 28 days. These results show mean and standard deviation of white blood cells (a), red blood cells (b), hematocrit (c), mean corpuscular volume (d), hemoglobin (e), platelets (f), mean corpuscular hemoglobin (g), mean corpuscular hemoglobin concentration (h). Bars represent mean ± standard deviation. Data is analyzed by T student test, and star represents significant difference from the control group ($p < 0.05$).

The ALT and AST are mainly distributed in liver cells, and their values rise with necrosis of liver cells. The levels of these enzymes correspond well with the extent of liver cell damage, and are commonly used as indicators of liver function. These two enzymes are distributed differently among the liver cells. The ALT mainly exists in the cytoplasm of liver cells, whereas the AST mainly exists in the cytoplasm and mitochondria of liver cells. Therefore, the increase of AST in our mice treated with the GSH-protected $Au_{25}$ NCs indicates acute damage to liver cells after 24 hours. However, biodistribution shows that the concentration of the GSH-protected $Au_{25}$ NCs in the liver is not as high as that reached by the BSA-protected $Au_{25}$ NCs, which indicates that the liver damage caused by the GSH-protected $Au_{25}$ NCs may be due to metabolism of the NCs. However, it is still not clear whether Au can be metabolized by liver. In contrast, the BSA-protected $Au_{25}$ NCs reach high concentration in the liver, but liver function is not significantly affected, showing that the BSA-protected $Au_{25}$ NCs don't cause direct damage to the liver. The CREA is another important indicator of kidney function. Endogenous human CREA is a product of muscle metabolism. In muscle, creatine mainly generates CREA slowly through nonenzymatic dehydration, which is then released into the blood, with excretion in the urine. The serum CREA concentration depends on the glomerular filtration rate. However, serum CREA is not entirely consistent with the CREA clearance rate, and the CREA clearance is more sensitive than serum CREA. In early renal dysfunction (decompensated), the CREA clearance rate and serum CREA are normal. When the glomerular filtration rate rises to above 50% of normal, serum CREA begins to rise rapidly. Therefore, when the serum CREA is significantly higher than normal, kidney function is serious damaged. The increase of CREA in mice treated with the two kinds of NCs is closely related to kidney function, although the kidney is not the main target organ for the GSH-protected $Au_{25}$ NCs. However, lots of $Au_{25}$ NCs can be found in urine and thus kidney participates in the metabolism of the GSH-protected $Au_{25}$ NCs. After 28 days, the GSH-protected $Au_{25}$ NCs don't induce any kidney damage, but the BSA-protected $Au_{25}$ NCs still cause some damage by decreasing total protein and globulin. It can be conceived that these toxicity responses of the BSA-protected $Au_{25}$ NCs are from the Au accumulation in spleen and liver. The $Au_{25}$ NCs firstly cause the acute immune response with decreasing thymus index. Subsequently, the continuous Au accumulation in liver and spleen destroys the immune system and causes the unrepairable damage. Further, the liver and kidney can not obtain the protectation by immune system. Finally, these damages can be reflected by the clinical hematology and biochemistry.

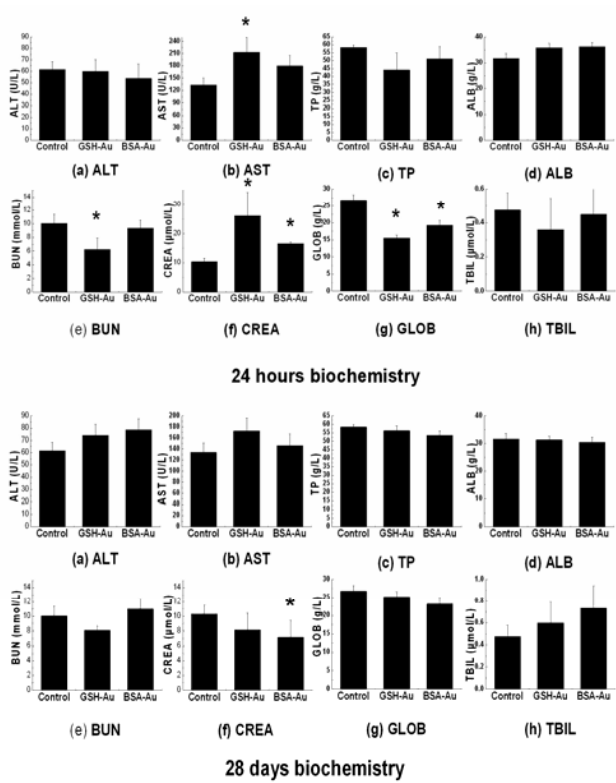

**Figure 7.** Blood biochemistry results of the GSH- and BSA-protected Au$_{25}$ NCs treated mice after 24 hours and 28 days. These results show mean and standard deviation of ALT (a), AST (b), total protein (c), albumin (d), blood urea nitrogen (e), creatinine (f), globulin (g), total bilirubin (h). Bars represent mean ± standard deviation. These results show the liver and kidney suffer serious damage for the two NCs after 24 hours. The liver and kidney function for the GSH-protected Au$_{25}$ NCs can be recovered after 28 days. Bars represent mean ± standard deviation. Data is analyzed by T student test, and star represents significant difference from the control group ($p < 0.05$).

*3.6. Outline of metabolism*

Synthetically, we summary these findings in Fig.8. The GSH-protected Au$_{25}$ NCs have slight aggregation and small NCs can be metabolized by urine, and thus these NCs have little accumulation in the liver. The BSA-protected Au$_{25}$ NCs, however, suffer significant aggregation in the blood and form large size NPs, and these NPs can be accumulated in liver and spleen without metabolism. Subsequently, the liver and kidney function are affected by injection of the BSA- and GSH-protected Au$_{25}$ NCs and 24 hours acute toxic response is observed at high concentration of 7550 μg/kg. However, the liver and kidney functions for GSH-protected Au$_{25}$ NCs can be recovered after 28 days treatment due to the clearance of Au. As a contrast, the toxicity of the BSA-protected Au$_{25}$ NCs can not be recovered due to the high accumulation of Au. In these clearance processing, kidney participate the metabolism of Au, and thus cause kidney damage at the short time, which is similar to the *in vivo* toxicity experiment of ZnS QDs [17]. The difference of biodistribution between the BSA- and GSH-protected Au$_{25}$ NCs treated mice shows that the surface chemistry of Au plays a very important role in biodistribution and toxicity. The extreme high Au content for the BSA-protected Au$_{25}$ NCs in liver can be utilized in targeting therapy for liver cancer. Meanwhile, the GSH-protected Au$_{25}$ NCs are easy to metabolize and have low content and accumulation in liver and spleen, and thus it has great potential and wide prospective for the *in vivo* imaging and therapy if the luminescence can be further enhanced in future.

## 4. Conclusions

*In vivo* renal clearance and toxicity studies of the BSA- and GSH-protected Au$_{25}$ NCs treated mice are carried out. Animal weight, hematology, biochemistry, and biodistributions are characterized at the concentration of 7550 μg/kg after 24 hours and 28 days. Biodistribution shows that the BSA-protected Au$_{25}$ NCs mainly be accumulated in the liver and spleen, and the GSH-protected Au$_{25}$ NCs have low concentration in all organs. After 24 hours, 36 % of the GSH-protected Au$_{25}$ NCs can be excreted by urine and only 1% of the BSA-protected Au$_{25}$ NCs can be excreted by urine. Further, 94% of Au in the GSH-protected Au$_{25}$ NCs can be metabolized by renal clearance, and less than 5% of Au can be metabolized in the BSA-protected Au$_{25}$ NCs after 28 days. For 24 hours toxicity, blood chemistry reveals that the WBC of the GSH- and BSA-protected Au$_{25}$ NCs treated mice has significantly increased. Further, biochemistry results show that the CREA of the GSH- and BSA-protected Au$_{25}$ NCs treated mice increase, and kidney function is affected. However, these toxicity responses of the GSH-protected Au$_{25}$ NCs can be eliminated after 28 days, but the BSA-protected Au$_{25}$ NCs still cause liver, kidney damage and the infection of mice,. The present work clearly shows the GSH-protected Au$_{25}$ NCs are metabolizable by renal clearance and have not induced the considerable toxicity responses. These conclusions are very important for future cancer therapy, drug delivery, and bioimaging of $Au_{25}$ NCs.

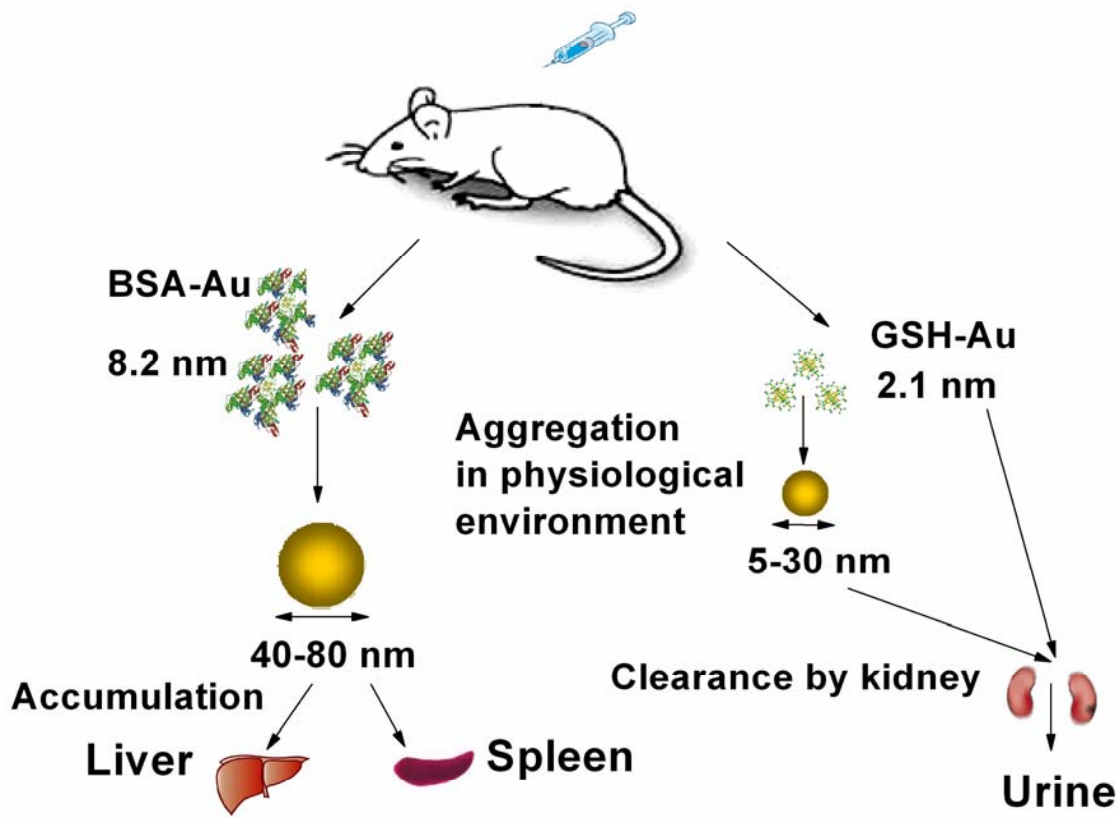

**Figure 8.** Outline of the biodistribution and renal clearance the GSH- and BSA-protected $Au_{25}$ NCs.

**Acknowledgements**
(This work was supported by Natural Science Foundation of China (Grant No. 81000668) and the Subject Development Foundation of Institute of Radiation Medicine, CAMS)